\documentclass[12pt,journal, onecolumn, draftcls]{IEEEtran}

\usepackage{verbatim}
\usepackage{amsmath,graphicx}
\usepackage{mathrsfs}
\usepackage{cite}
\usepackage{amsthm}
\usepackage{epsfig,psfrag}
\usepackage{amsfonts,amssymb}

\begin{document}
\title{Finite-Length and Asymptotic Analysis of Correlogram for Undersampled Data}

\author{Mahdi~Shaghaghi and~Sergiy~A.~Vorobyov
\thanks{M. Shaghaghi and S. A. Vorobyov are with the Department
of Electrical and Computer Engineering, University of Alberta,
Edmonton, AB, T6G 2V4 Canada (e-mail: mahdi.shaghaghi@ualberta.ca;
svorobyo@ualberta.ca). S.~A.~Vorobyov is currently on leave and he
is with Aalto University, Department of Signal Processing and
Acoustics, Finland. S.~A.~Vorobyov is the corresponding author.}
\thanks{Some results of this work have been reported in ICASSP'12, Kyoto, Japan.}
}

%

\maketitle

\begin{abstract}
This paper studies a spectrum estimation method for the case that
the samples are obtained at a rate lower than the Nyquist rate. The
method is referred to as the correlogram for undersampled data. The
algorithm partitions the spectrum into a number of segments and
estimates the average power within each spectral segment. This
method is able to estimate the power spectrum density of a signal
from undersampled data without essentially requiring the signal to
be sparse. We derive the bias and the variance of the spectrum
estimator, and show that there is a tradeoff between the accuracy of
the estimation, the frequency resolution, and the complexity of the
estimator. A closed-form approximation of the estimation variance is
also derived, which clearly shows how the variance is related to
different parameters. The asymptotic behavior of the estimator is
also investigated, and it is proved that this spectrum estimator is
consistent. Moreover, the estimation made for different spectral
segments becomes uncorrelated as the signal length tends to
infinity. Finally, numerical examples and simulation results are
provided, which approve the theoretical conclusions.

\end{abstract}

\begin{IEEEkeywords}
Spectral analysis, correlogram, undersampling, consistency.
\end{IEEEkeywords}

\IEEEpeerreviewmaketitle

\section{Introduction}
Spectrum estimation from a finite set of noisy measurements is a
classical problem with wide applications in communications,
astronomy, seismology, radar, sonar signal processing, etc.
\cite{VanTrees02:EST}, \cite{Manolakis00:EST}. Classical methods
such as the periodogram, the correlogram, the multiple signal
classification (MUSIC) method \cite{Schmidt86:MUSIC}, and the
estimation of signal parameters via rotational invariance techniques
(ESPRIT) \cite{Roy89:ESPRIT} estimate the spectrum based on the
\textit{Nyquist samples} (samples obtained at the Nyquist rate). In
practice, the rate at which the measurements are collected can be
restricted. Examples include the case when the speed of the sampling
hardware is limited or the case when samples of a data record are
missing. Therefore, it is desirable to make spectrum estimation from
measurements obtained at a rate lower than the Nyquist rate.

In \cite{Vaidyanathan90:polyphase} and \cite{Herley99:min_rate},
authors have studied signal reconstruction from sub-Nyquist samples
which are obtained by nonuniform sampling. The methods in these
works consider band-limited and multi-band signals with the prior
knowledge of the spectral support of the signal, i.e., the position
of the frequency bands. In \cite{Eldar09:blind_CS}, algorithms for
signal recovery from undersampled data without the prior knowledge
of the spectral support except for the number and the widths of the
frequency bands have been proposed. The methods in
\cite{Vaidyanathan90:polyphase}--\hspace{-0.05mm}\cite{Eldar09:blind_CS}
aim at reconstructing the signal, whereas depending on the
application, e.g., cognitive radio systems \cite{Haykin05:cog_rad},
one might be only interested in recovering the spectral information
of the signal. In \cite{Gilbert08:Sparse_Fourier}, authors have
shown that for signals with sparse Fourier representations, i.e.,
signals which have only a few nonzero coefficients in the Fourier
basis, the Fourier coefficients can be estimated using a subset of
the Nyquist samples. In \cite{Tian07:cognitive_CS}, power spectral
density (PSD) estimation based on \textit{compressive sensing} (CS)
techniques \cite{Donoho06:CS}, \cite{Candes06:ExactCS} with
applications in wideband cognitive radios has been introduced. In
\cite{Baraniuk10:Spectral_CS} and \cite{MSH11:Nested_LS}, the
possibility of recovering signals sparse in the discrete-time
Fourier transform (DTFT) domain from compressive samples obtained at
a rate lower than the Nyquist rate has been demonstrated. In
\cite{Candes12:CS_PSD_arXiv} and \cite{Tang12:off_grid}, the
super-resolution problem has been addressed where the position of a
few sparse sources is resolved with infinite precision from only
samples of the low-frequency end of the spectrum. In the
super-resolution methods, the information of the high-frequency
portion of the spectrum is extrapolated based on the samples of the
low-frequency part. However, this is only possible for sparse
sources with the additional constraint that the distance between any
two sources be larger than a minimum value, i.e., the sources be
well-separated.

For all of the above mentioned methods, the sparsity of the signal
is a requirement for successful recovery of the spectrum. In
\cite{Lexa11:CS_PSD_arXiv}, PSD estimation from a subset of the
Nyquist samples has been considered . The introduced method is able
to estimate the PSD from undersampled data without essentially
requiring the signal to be sparse. We will show in this paper that
this is achieved with a trade-off between the spectral resolution
and the estimation accuracy. We refer to this method as the
\textit{correlogram for undersampled data}. In this method, samples
are collected using multiple channels, each operating at a rate $L$
times lower than the Nyquist rate. This method of sampling is known
as the \textit{multi-coset sampling}
\cite{FengBresler96:multi-coset}. The correlogram for undersampled
data partitions the spectrum into $L$ segments (subbands), and it
estimates the average power within each spectral segment. The
frequency resolution of the estimator is given by the width of each
spectral segment. In this paper, we equivalently use the number of
spectral segments $L$ as the frequency resolution of the estimator
(with larger $L$ meaning higher resolution or narrower segments). In
\cite{Ariananda12:psd}, PSD estimation based on sub-Nyquist samples
is also considered. The main difference to
\cite{Lexa11:CS_PSD_arXiv}, however, is that in
\cite{Ariananda12:psd}, the introduced method estimates samples of
the PSD, whereas in the correlogram for undersampled data, the
average power within subbands is estimated. As a result, the
correlogram for undersampled data is less computationally complex
\cite{Lexa11:CS_PSD_arXiv}.

The advantage of the correlogram for undersampled data as mentioned
above is its ability in estimating the PSD from sub-Nyquist samples
without necessarily imposing sparsity conditions on the signal. This
is not, however, achieved without paying a price, and it is,
therefore, of significant importance to know the associated
tradeoffs. The focus of this paper is to analyze the performance of
the correlogram for undersampled data and to formulate the
associated tradeoffs.

We first study the correlogram for undersampled data by computing
the bias of the estimator. Next, the covariance matrix of the
estimator is derived, and using our derivations, we show that for
finite-length signals, there exists a tradeoff between the
estimation accuracy, the frequency resolution, and the complexity of
the estimator.\footnote{Note that the complexity is a critical issue
in a number of applications, for example, for fighting the curse of
dimensionality for data acquisition in exploration seismology
\cite{Herrmann}.} For the case of a white Gaussian process, we
derive a closed-form expression for the estimation variance, which
clearly shows how the variance is related to different parameters.
Moreover, we prove that the estimation bias and variance tend to
zero asymptotically. Therefore, the correlogram for undersampled
data is a consistent estimator. This is in contrast with the
conventional correlogram which does not enjoy the consistency
property \cite{Hayes96:sig_proc}. Besides, we show that similar to
the conventional correlogram, the correlogram for undersampled data
makes uncorrelated estimations for different spectral segments as
the signal length goes to infinity.

The rest of the paper is organized as follows. The correlogram for
undersampled data is revised in Section \ref{sec:undr_corr}.
Specifically, we introduce a practical implementation of the filters
used in the estimator. In Section \ref{sec:bias_var}, the bias and
the covariance matrix of the correlogram for undersampled data are
derived, and a closed-form expression for the estimation variance is
given. Section \ref{sec:examples_sim} presents some numerical
examples on the estimation bias and variance of the correlogram
method for finite-length signals. Finally, Section
\ref{sec:conclude} concludes the paper. The proofs and derivations
are given in Appendices. This paper is reproducible research and the
software needed to generate the numerical results will be provided
to the IEEE Xplore together with the paper.

\section{Correlogram for Undersampled Data}\label{sec:undr_corr}
Consider a wide-sense stationary (WSS) stochastic process $x(t)$
bandlimited to $W/2$ Hz with power spectral density (PSD) $P_x(f)$.
Let $x(t)$ be sampled using the multi-coset (MC) sampler as
described in \cite{Lexa11:CS_PSD_arXiv}. Samples are collected by a
multi-channel system. The $i$-th channel ($1\leq i\leq q$) samples
$x(t)$ at the time instants $t=(nL+c_i)T$ for $n=0,~1,~2,~\ldots$,
where $T$ is the Nyquist period ($T=1/W$), $L$ is a suitable
positive integer, and $q<L$ is the number of sampling channels. The
time offsets $c_i$ ($1\leq i\leq q$) are distinct non-negative
integer numbers less than $L$, and the set $\{c_i\}$ is referred to
as the \textit{sampling pattern}. Let the output of the $i$-th
channel be denoted by $y_i(n)=x\left((nL+c_i)T\right)$. The $i$-th
channel can be implemented by a system that shifts $x(t)$ by $c_iT$
seconds and then samples uniformly at a rate of $1/(LT)$ Hz. The
samples obtained in this manner form a subset of the Nyquist
samples. The average sampling rate is $q/(LT)$ Hz, and it is less
than the Nyquist rate since $q<L$.

Given the MC samples, the first step of the correlogram for
undersampled data method is to undo the time shift that each channel
imposes on the signal. Let $z_i(n)$ be defined as $y_i(n)$ delayed
by a fractional delay equal to $c_i/L$. Let also $a$ and $b$ denote
two channel indices. It is shown in \cite{Lexa11:CS_PSD_arXiv} that
the cross-correlation function $r_{z_a z_b}(k) =
E\{z_a(n+k)z_b^*(n)\}$ at $k=0$ is given by
\begin{equation}
r_{z_a z_b}(0) = \sum_{l=1}^L e^{-j\frac{2\pi}{L}(c_a-c_b)m_l}
P_x(m_l)\label{eq:base_cov_elmnt}
\end{equation}
where $E\{\cdot\}$ stands for the expectation operator, $L$ is an
odd number, $m_l = -\frac{1}{2}(L+1)+l$, and $P_x(m_l)$ is defined
as
\begin{equation}
P_x(m_l) \triangleq \int_{-\frac{W}{2L}}^{\frac{W}{2L}} P_x\left(f -
 \frac{W}{L}m_l\right) df.
\end{equation}
Consider partitioning the bandwidth of $x(t)$ into $L$ equal
segments. Then, for a given $m_l$, $\frac{L}{W}P_x(m_l)$ is equal to
the average power of the process $x(t)$ within the spectral segment
$\big[\frac{W}{2} - \frac{W}{L}l,$ $\frac{W}{2} -
\frac{W}{L}(l-1)\big)$.

Let us arrange the elements of the cross-correlation function
$r_{z_a z_b}(0)$ ($1\leq a,b \leq q$) in a matrix $\boldsymbol{R}_z
\in \mathbb{C}^{q \times q}$ such that $[\boldsymbol{R}_{z}]_{a,b} =
r_{z_a z_b}(0)$. Note that $\boldsymbol{R}_{z}$ is a Hermitian
matrix with equal diagonal elements. Then, it is sufficient to let
the indices $a$ and $b$ just refer to the elements of the upper
triangle and the first diagonal element of $\boldsymbol{R}_{z}$.
Therefore, there are $Q = q(q-1)/2+1$ equations of type
\eqref{eq:base_cov_elmnt}. In matrix-vector form,
\eqref{eq:base_cov_elmnt} can be rewritten as
\begin{equation}
\boldsymbol{u}=\boldsymbol{\Psi}\boldsymbol{v}\label{eq:uPsiv}
\end{equation}
where $\boldsymbol{v}=[v_1,~v_2,~\ldots,~v_L]^{T}\in
\mathbb{R}^{L\times 1}$ consists of the elements $v_l = P_x(m_l)$,
$(\cdot)^T$ stands for the transposition operator,
$\boldsymbol{u}=[u_1,~u_2,~\ldots,~u_Q]^{T} \in \mathbb{C}^{Q\times
1}$ is composed of $u_1=[\boldsymbol{R}_{z}]_{1,1}$ and
$u_2,\ldots,~u_Q$ corresponding to the elements of the upper
triangle of $\boldsymbol{R}_{z}$, and $\boldsymbol{\Psi} \in
\mathbb{C}^{Q\times L}$ consists of the elements given by
\begin{equation}
[\boldsymbol{\Psi}]_{k,l}=e^{-j \omega_k m_l}\label{eq:Psi_elements}
\end{equation}
where $\omega_k = \frac{2\pi}{L}(c_a-c_b)$, ($1\leq l \leq L$ and
$1\leq k\leq Q$). Note that $a$ and $b$ are obtained from $k$ based
on the arrangement of the elements of $\boldsymbol{R}_{z}$ in
$\boldsymbol{u}$.

Since the elements of $\boldsymbol{v}$ are real-valued, the number
of equations in \eqref{eq:uPsiv} can be doubled\footnote{Doubling
the number of equations is beneficial in turning an underdetermined
system of equations into an overdetermined system.} by solving
$\breve{\boldsymbol{u}}=\breve{\boldsymbol{\Psi}}\boldsymbol{v}$,
where
$\breve{\boldsymbol{u}}\triangleq[Re(\boldsymbol{u}),~Im(\boldsymbol{u})]^{T}
\in \mathbb{R}^{2Q\times 1}$ and
$\breve{\boldsymbol{\Psi}}\triangleq[Re(\boldsymbol{\Psi}),~Im(\boldsymbol{\Psi})]^{T}
\in \mathbb{R}^{2Q\times L}$.

Suppose $\breve{\boldsymbol{\Psi}}$ is full rank and $2Q\geq L$.
Then,
$\breve{\boldsymbol{u}}=\breve{\boldsymbol{\Psi}}\boldsymbol{v}$ is
an overdetermined system and $\boldsymbol{v}$ can be obtained using
the pseudoinverse of $\breve{\boldsymbol{\Psi}}$ as
\begin{equation}
\boldsymbol{v}=(\breve{\boldsymbol{\Psi}}^T\breve{\boldsymbol{\Psi}})^{-1}\breve{\boldsymbol{\Psi}}^T\breve{\boldsymbol{u}}.\label{eq:Psduinv_est}
\end{equation}

The cross-correlation function $r_{z_a z_b}(k)$ can be estimated
from a finite number of samples as
\begin{equation}
\widehat{r}_{z_a z_b}(k) =
\frac{1}{N}\sum_{n=0}^{N-|k|-1}\widehat{z}_a(n+k) \widehat{z}_b(n)
\end{equation}
where $N$ is the number of samples obtained from each channel, and
$\widehat{z}_a(n+k)$ and $\widehat{z}_b(n)$ are obtained by delaying
$y_a(n+k)$ and $y_b(n)$ for $c_a/L$ and $c_b/L$ fractions,
respectively. Next, the elements of the matrix $\boldsymbol{R}_{z}$
are estimated as
\begin{equation}
[\widehat{\boldsymbol{R}}_z]_{a,b} = \widehat{r}_{z_a z_b}(0) =
\frac{1}{N}\sum_{n=0}^{N-1}\widehat{z}_a(n)
\widehat{z}_b^*(n).\label{eq:est_Rz}
\end{equation}
The fractional delays $c_a/L$ and $c_b/L$ can be implemented by
fractional delay (FD) filters. In \cite{Lexa11:CS_PSD_arXiv},
authors consider using ideal FD filters which have infinite impulse
responses. Then, for the purpose of implementation, these filters
are truncated using a rectangular window whose width is twice the
signal length $N$. Consequently, the length of the filters can be
quite large as $N$ increases. Here, we consider using causal finite
impulse response (FIR) filters which have two practical advantages
\cite{MSH12:Corr}: first, the length of the filters are fixed, and
second, they enjoy causality. As for the analysis, we will use a
general formulation for the FIR FD filters, and for numerical
examples, we will use the Lagrange interpolator
\cite{Laakso96:Delay_filter}.

FIR FD filters perform the best when the total delay is
approximately equal to half of the order of the filter
\cite{Laakso00:Delay_filter}. The fractional delays $c_a/L$ and
$c_b/L$ are positive numbers less than one, and the performance of
the FIR FD filters is very poor with such delays. To remedy this
problem, a suitable integer delay can be added to the fractional
part. Note that $\widehat{r}_{z_a z_b}(k)$ is the inverse
discrete-time Fourier transform (DTFT) of $(1/N)
\widehat{Z}_a\left(e^{j2\pi f L/W}\right)
\widehat{Z}_b^*\left(e^{j2\pi f L/W}\right)$, where
$\widehat{Z}_a\left(e^{j2\pi f L/W}\right)$ and
$\widehat{Z}_b\left(e^{j2\pi f L/W}\right)$ are the DTFT of
$\widehat{z}_a(n)$ and $\widehat{z}_b(n)$, respectively
\cite{Johnson93:DTFT}. Then, considering that
\begin{eqnarray}
&& \hspace{-10mm} \widehat{Z}_a\left(e^{j2\pi f\frac{L}{W}}\right)
\widehat{Z}_b^*\left(e^{j2\pi f\frac{L}{W}}\right) = \nonumber\\
&& \hspace{-8mm} \left[\widehat{Z}_a\left(e^{j2\pi
f\frac{L}{W}}\right)e^{-jD2\pi f\frac{L}{W}}\right]
\left[\widehat{Z}_b\left(e^{j2\pi f\frac{L}{W}}\right)e^{-jD2\pi
f\frac{L}{W}}\right]^*
\end{eqnarray}
we can rewrite \eqref{eq:est_Rz} as
\begin{equation}
[\widehat{\boldsymbol{R}}_z]_{a,b} = \frac{1}{N} \sum_{n=0}^{N-1}
\widehat{z}_a(n-D) \widehat{z}_b^*(n-D)\label{eq:est_Rz_D}
\end{equation}
where $D$ is a suitable integer number close to half of the order of
the FD filter.

Let $h_a(n)$ be the impulse response of a causal filter that delays
a signal for $c_a/L+D$. Furthermore, let us assume that the length
of $h_a(n)$ is large enough, so that its deviation from an ideal FD
filter can be ignored. Therefore, $z_a(n-D)$ can be written as
\begin{equation}
z_a(n-D) = \sum_{r=0}^{N_h-1} h_a(r) y_a(n-r) \label{eq:z_aD}
\end{equation}
where $N_h$ is the length of the filter's impulse response. For a
limited number of samples, we have
\begin{eqnarray}
\widehat{z}_a(n-D) \hspace{-2mm} &=& \hspace{-2mm} \sum_{r=0}^{N_h-1} h_a(r) y_a(n-r) W_D(n-r) \nonumber\\
&=& \hspace{-2mm} \sum_{r = n - N_h + 1}^{n} h_a(n-r) y_a(r)
W_D(r)\label{eq:est_z}
\end{eqnarray}
where $W_D(n)$ is a window of length $N$ which equals $1$ for $0
\leq n \leq N-1$ and is equal to zero elsewhere. Using the elements
of $\widehat{\boldsymbol{R}}_z$, the vector
$\widehat{\breve{\boldsymbol{u}}}$ is formed as an estimation for
$\breve{\boldsymbol{u}}$. Next, $\widehat{\boldsymbol{v}}$ (the
estimation for $\boldsymbol{v}$) is formed by replacing
$\breve{\boldsymbol{u}}$ with $\widehat{\breve{\boldsymbol{u}}}$ in
\eqref{eq:Psduinv_est} as
\begin{equation}
\widehat{\boldsymbol{v}}=(\breve{\boldsymbol{\Psi}}^T\breve{\boldsymbol{\Psi}})^{-1}\breve{\boldsymbol{\Psi}}^T\widehat{\breve{\boldsymbol{u}}}.\label{eq:Psduinv_est2}
\end{equation}
Finally, let us define $\widehat{\boldsymbol{p}} \in
\mathbb{R}^{L\times 1}$ as
\begin{equation}
\widehat{\boldsymbol{p}}\triangleq
\frac{L}{W}\widehat{\boldsymbol{v}}. \label{eq:p_hat_def}
\end{equation}
The elements of $\widehat{\boldsymbol{p}}$ give an estimation for
the average power within each spectral segment.

\section{Bias and Variance of Correlogram for Undersampled
Data}\label{sec:bias_var}

Consider a Gaussian WSS signal $x(t)$ bandlimited to $W/2$ Hz, and
let $x(m)$ be the samples of the signal obtained at the Nyquist rate
($m \in \mathbb{Z}$). Let also $r_x(k)=E\{x(m+k)x^*(m)\}$ and
$P_x(e^{j 2\pi f/W}) = \text{DTFT} \left\{r_x(k)\right\}$ be the
autocorrelation function and the PSD of $x(m)$, respectively.
Furthermore, consider a zero-mean Gaussian random process $e(t)$
bandlimited to $W/2$ Hz with a flat PSD $P_e(f) = \sigma^2 / W$. The
autocorrelation function of $e(t)$ is $r_e(\tau) = \sigma^2
\text{sinc}(W\tau)$. Let $e(m)$ be the samples of $e(t)$ obtained at
the Nyquist rate. Then, the autocorrelation function of $e(m)$ is
given by
\begin{equation}
r_e(k) = \sigma^2 \text{sinc}(W k/W) = \sigma^2 \delta(k)
\end{equation}
where $\delta(k)$ is the Kronecker delta. Therefore, the PSD of
$e(m)$ is given by $P_e(e^{j 2\pi f/W}) = \sigma^2$.   Now, consider
a filter $h_x(m)$ such that $\sigma^2 |H_x(e^{j 2\pi f/W})|^2$ is
equal to $P_x(e^{j 2\pi f/W})$, where $H_x(e^{j 2\pi f/W})$ is the
DTFT of $h_x(m)$. Therefore, we have
\begin{equation}
P_x(e^{j 2\pi f/W}) = |H_x(e^{j 2\pi f/W})|^2 P_e(e^{j 2\pi f/W}).
\end{equation}
As a result, $x(m)$ can be considered as $e(m)$ filtered by $h_x(m)$
since the output of the filter has the same PSD as $P_x(e^{j 2\pi
f/W})$. Then, the output of the $i$-th sampling channel can be
written as
\begin{equation}
y_i(n) = x(nL + c_i) = \sum_{m \in \mathbb{Z}} h_x(m) e(nL + c_i -
m). \label{eq:ith_chl_fltrd}
\end{equation}
Let $a$ and $b$ denote two channel indices. The cross-correlation
function $r_{y_a y_b}(k) = E\{y_a(n+k)y_b^*(n)\}$ is given by
\begin{eqnarray}
\hspace{-1mm} r_{y_a y_b}(k) \hspace{-2mm} &=& \hspace{-2mm} \sum_{m
\in \mathbb{Z}}
\sum_{l \in \mathbb{Z}} h_x(m) h_x^*(l) \nonumber\\
&&  E\left\{ e((n+k)L + c_a - m) e^*(nL + c_b - l) \right\}
\nonumber\\
&=& \hspace{-2mm} \sum_{m \in \mathbb{Z}} \sum_{l \in \mathbb{Z}}
h_x(m) h_x^*(l) r_e(kL + l - m + c_a - c_b) \nonumber\\
&=& \hspace{-2mm} \sigma^2 \sum_{m \in \mathbb{Z}} h_x(m) h_x^*(-kL
+ m + c_b - c_a).\label{eq:r_yayb_e}
\end{eqnarray}
Furthermore, using \eqref{eq:z_aD},
$[\boldsymbol{R}_{\boldsymbol{z}}]_{a,b}$ can be written as
\begin{eqnarray}
[\boldsymbol{R}_{\boldsymbol{z}}]_{a,b} \hspace{-2mm} &=&
\hspace{-2mm} E\{z_a(n-D) z_b^*(n-D)\} \nonumber\\
&=& \hspace{-2mm} \sum_{r=0}^{N_h-1} \sum_{p=0}^{N_h-1} h_a(r)
h_b(p) r_{y_a y_b}(p-r).\label{eq:Rz_ab_D}
\end{eqnarray}

\subsection{Bias Analysis}
The bias of the correlogram for undersampled data estimator is given
by
\begin{equation}
E\{\widehat{\boldsymbol{p}}\} - \boldsymbol{p} = \frac{L}{W} \left(
E\{\widehat{\boldsymbol{v}}\} - \boldsymbol{v}
\right)\label{eq:bias_p}
\end{equation}
where $\boldsymbol{p} = (L/W) \boldsymbol{v}$. The expected value of
$\widehat{\boldsymbol{v}}$ is obtained using \eqref{eq:Psduinv_est2}
as
\begin{equation}
E\{\widehat{\boldsymbol{v}}\}=(\breve{\boldsymbol{\Psi}}^T\breve{\boldsymbol{\Psi}})^{-1}\breve{\boldsymbol{\Psi}}^T
E\{\widehat{\breve{\boldsymbol{u}}}\}. \label{eq:Ev1}
\end{equation}
Computing $E\{\widehat{\breve{\boldsymbol{u}}}\}$ requires finding
the expected value of the real and imaginary parts of
$\widehat{\boldsymbol{R}}_{\boldsymbol{z}}$. The expectation
operation can be performed before taking the real or imaginary parts
of $\widehat{\boldsymbol{R}}_{\boldsymbol{z}}$, as these operators
are linear. Moreover, \eqref{eq:est_Rz_D} is used to form
$\widehat{\boldsymbol{R}}_{\boldsymbol{z}}$. Taking expectation from
both sides of \eqref{eq:est_Rz_D} along with using \eqref{eq:est_z}
results in
\begin{eqnarray}
\hspace{-0mm}E\{[\widehat{\boldsymbol{R}}_{\boldsymbol{z}}]_{a,b}\}\hspace{-2mm}&=&\hspace{-2mm}
\frac{1}{N} \sum_{n=0}^{N-1} \sum_{r=0}^{N_h-1} \sum_{p=0}^{N_h-1}
\nonumber \\
&&\hspace{-21mm}h_a(r) h_b(p) W_D(n-r) W_D(n-p) E\{y_a(n-r)
y_b^*(n-p)\} \nonumber \\
&=&\hspace{-2mm} \sum_{r=0}^{N_h-1} \sum_{p=0}^{N_h-1} h_a(r) h_b(p)
r_{y_a y_b}(p-r) \times
\nonumber \\
&&\hspace{10mm}\frac{1}{N} \sum_{n=0}^{N-1} W_D(n-r) W_D(n-p)
\label{eq:ERz0}
\end{eqnarray}
With the assumption that the number of samples $N$ is larger than
the length of the fractional delay filters $N_h$, the last summation
of \eqref{eq:ERz0} can be simplified to
\begin{equation}
\sum_{n=0}^{N-1} W_D(n-r) W_D(n-p) = N - \text{max}(r,p).
\end{equation}
Therefore, \eqref{eq:ERz0} can be rewritten as
\begin{eqnarray}
\hspace{-0mm}E\{[\widehat{\boldsymbol{R}}_{\boldsymbol{z}}]_{a,b}\}\hspace{-2mm}&=&\hspace{-2mm}
[\boldsymbol{R}_{\boldsymbol{z}}]_{a,b} - \frac{1}{N}
\sum_{r=0}^{N_h-1} \sum_{p=0}^{N_h-1}
\nonumber \\
&&\hspace{-0mm} h_a(r) h_b(p) r_{y_a y_b}(p-r)\text{max}(r,p)
\label{eq:ERz01}
\end{eqnarray}
where $[\boldsymbol{R}_{\boldsymbol{z}}]_{a,b}$ is given by
\eqref{eq:Rz_ab_D}.

It can be seen from \eqref{eq:ERz01} that as $N$ tends to infinity,
$E\{[\widehat{\boldsymbol{R}}_{\boldsymbol{z}}]_{a,b}\}$ tends to
$[\boldsymbol{R}_{\boldsymbol{z}}]_{a,b}$. Therefore,
$\widehat{\boldsymbol{R}}_z$ is an asymptotically unbiased estimator
of $\boldsymbol{R}_{z}$. Since $\widehat{\breve{\boldsymbol{u}}}$
consists of the elements of $\widehat{\boldsymbol{R}}_z$ and the
operation of taking the real and imaginary parts are linear, it
follows that $\widehat{\breve{\boldsymbol{u}}}$ is also an
asymptotically unbiased estimator of $\breve{\boldsymbol{u}}$.
Furthermore, letting the number of samples tend to infinity in
\eqref{eq:Ev1} and using \eqref{eq:Psduinv_est}, we find that
\begin{eqnarray}
\lim_{N\rightarrow\infty}
E\{\widehat{\boldsymbol{v}}\}\hspace{-2mm}&=&\hspace{-2mm}(\breve{\boldsymbol{\Psi}}^T\breve{\boldsymbol{\Psi}})^{-1}\breve{\boldsymbol{\Psi}}^T\lim_{N\rightarrow\infty}E\{\widehat{\breve{\boldsymbol{u}}}\}\nonumber\\
\hspace{-2mm}&=&\hspace{-2mm}(\breve{\boldsymbol{\Psi}}^T\breve{\boldsymbol{\Psi}})^{-1}\breve{\boldsymbol{\Psi}}^T\breve{\boldsymbol{u}}=\boldsymbol{v}.
\end{eqnarray}
In other words, $\widehat{\boldsymbol{v}}$ is also an asymptotically
unbiased estimator of $\boldsymbol{v}$. Finally, it can be concluded
from \eqref{eq:bias_p} that the correlogram for undersampled data
estimator $\widehat{\boldsymbol{p}}$ is asymptotically unbiased.

Next, we consider the case that the input signal $x(t)$ is equal to
the white Gaussian random process $e(t)$. It is shown in Appendix
\ref{sec:appndx_bias} that
\begin{equation}
E\{\widehat{\boldsymbol{p}}\} = H_1 \boldsymbol{p} =
H_1\frac{\sigma^2}{W}\boldsymbol{1}_L \label{eq:Ephat}
\end{equation}
where $\boldsymbol{1}_L$ is the column vector of length $L$ with all
its elements equal to $1$, and $H_1$ is given by
\begin{equation}
H_1=\frac{1}{N}\sum_{r=0}^{N_h-1}(N-r)h_1^2(r).\label{eq:H1_def}
\end{equation}
Therefore, the bias of the correlogram for undersampled data
estimator in this case is given by
\begin{equation}
E\{\widehat{\boldsymbol{p}}\} - \boldsymbol{p} =
(1-H_1)\frac{\sigma^2}{W}\boldsymbol{1}_L.
\label{eq:bias_p_white_closed}
\end{equation}

\subsection{Variance Analysis}\label{sec:var}
The covariance matrix of the correlogram for undersampled data is
given by
\begin{eqnarray}
\mathcal{C}_{\widehat{\boldsymbol{p}}} \hspace{-2mm} &=&
\hspace{-2mm} E\left\{\left(\widehat{\boldsymbol{p}}-
E\{\widehat{\boldsymbol{p}}\}\right)\left(\widehat{\boldsymbol{p}}-
E\{\widehat{\boldsymbol{p}}\}\right)^T\right\} \nonumber\\
&=& \hspace{-2mm}
E\{\widehat{\boldsymbol{p}}\widehat{\boldsymbol{p}}^T\}-
E\{\widehat{\boldsymbol{p}}\} {E\{\widehat{\boldsymbol{p}}\}}^T.
\label{eq:cov_p_def}
\end{eqnarray}
The diagonal elements of $\mathcal{C}_{\widehat{\boldsymbol{p}}}$
are the estimation variance of each spectral segment. The
off-diagonal elements of $\mathcal{C}_{\widehat{\boldsymbol{p}}}$
represent the correlation between pairs of the estimations made for
different spectral segments.

It follows from \eqref{eq:Psduinv_est2} and \eqref{eq:p_hat_def}
that
\begin{equation}
E\{\widehat{\boldsymbol{p}}\widehat{\boldsymbol{p}}^T\}=\left(\frac{L}{W}\right)^2
(\breve{\boldsymbol{\Psi}}^T\breve{\boldsymbol{\Psi}})^{-1}\breve{\boldsymbol{\Psi}}^T\boldsymbol{U}
\breve{\boldsymbol{\Psi}}(\breve{\boldsymbol{\Psi}}^T\breve{\boldsymbol{\Psi}})^{-1}\label{eq:EppT}
\end{equation}
where $\boldsymbol{U}\triangleq
E\{\widehat{\breve{\boldsymbol{u}}}\widehat{\breve{\boldsymbol{u}}}^T\}
\in \mathbb{R}^{2Q \times 2Q}$. Computation of the elements of
$\boldsymbol{U}$ involves taking expectation of the multiplication
of the real or imaginary parts of the elements of
$\widehat{\boldsymbol{R}}_z$. We will use the following lemma
\cite{Stoica89:ML_CRB} for interchanging the expectation and the
operation of taking real or imaginary parts.

\theoremstyle{plain}
\newtheorem{lemma_xy}{Lemma}[]
\begin{lemma_xy}
Let $x$ and $y$ be two arbitrary complex numbers. The following
equations hold
\begin{eqnarray}
Re(x)Re(y)\hspace{-2mm}&=&\hspace{-2mm}\frac{1}{2}\left(Re(xy)+Re(xy^*)\right)\label{eq:RexRey}\\
Im(x)Im(y)\hspace{-2mm}&=&\hspace{-2mm}-\frac{1}{2}\left(Re(xy)-Re(xy^*)\right)\label{eq:ImxImy}\\
Re(x)Im(y)\hspace{-2mm}&=&\hspace{-2mm}\frac{1}{2}\left(Im(xy)-Im(xy^*)\right).\label{eq:RexImy}
\end{eqnarray}
\end{lemma_xy}

The elements of $\boldsymbol{U}$ can be easily obtained using
$E\{[\widehat{\boldsymbol{R}}_z]_{a,b}[\widehat{\boldsymbol{R}}_z]_{c,d}\}$,
$E\{[\widehat{\boldsymbol{R}}_z]_{a,b}[\widehat{\boldsymbol{R}}_z]_{c,d}^*\}$,
and Lemma $1$, where $[\widehat{\boldsymbol{R}}_z]_{a,b}$ and
$[\widehat{\boldsymbol{R}}_z]_{c,d}$ are the elements of
$\widehat{\boldsymbol{R}}_z$ used for forming
$\widehat{\breve{\boldsymbol{u}}}$. Let the outputs of the sampling
channels be given by \eqref{eq:ith_chl_fltrd}. Using
\eqref{eq:est_Rz_D} and \eqref{eq:est_z}, we obtain
\begin{eqnarray}
&& \hspace{-7mm} E\{[\widehat{\boldsymbol{R}}_z]_{a,b}[\widehat{\boldsymbol{R}}_z]_{c,d}\} = \nonumber\\
&& \hspace{-3mm} \frac{1}{N^2} \sum_{n=0}^{N-1} \hspace{2mm}
\sum_{\underset{(n-N_h+1)}{r=}}^n \sum_{\underset{(n-N_h+1)}{p=}}^n
\hspace{1mm} \sum_{u=0}^{N-1} \hspace{2mm}
\sum_{\underset{(u-N_h+1)}{s=}}^u \sum_{\underset{(u-N_h+1)}{m=}}^u \nonumber\\
&& \qquad h_a(n-r)h_b(n-p)h_c(u-s)h_d(u-m)\times\nonumber\\
&& \qquad W_D(r) W_D(p) W_D(s) W_D(m) \times\nonumber\\
&& \qquad E\{y_a(r)y_b^*(p)y_c(s)y_d^*(m)\} = \nonumber\\
&& \hspace{-3mm} \frac{1}{N^2} \sum_{n=0}^{N-1} \hspace{2mm}
\sum_{\underset{(0,n-N_h+1)}{r=\max}}^n
\sum_{\underset{(0,n-N_h+1)}{p=\max}}^n \hspace{1mm}
\sum_{u=0}^{N-1} \hspace{2mm}
\sum_{\underset{(0,u-N_h+1)}{s=\max}}^u \sum_{\underset{(0,u-N_h+1)}{m=\max}}^u \nonumber\\
&& \hspace{-1.5mm} h_a(n-r)h_b(n-p)h_c(u-s)h_d(u-m)\times\nonumber\\
&& \hspace{-1.5mm} \left(r_{y_a y_b}(r-p) r_{y_c y_d}(s-m) +
r_{y_a y_d}(r-m) r_{y_c y_b}(s-p)\right). \nonumber\\
\label{eq:ERzRz1}
\end{eqnarray}
The last line in \eqref{eq:ERzRz1} is obtained using the forth-order
moment of Gaussian random processes.

In a similar way,
$E\{[\widehat{\boldsymbol{R}}_z]_{a,b}[\widehat{\boldsymbol{R}}_z]_{c,d}^*\}$
can be obtained as
\begin{eqnarray}
&& \hspace{-7mm} E\{[\widehat{\boldsymbol{R}}_z]_{a,b}[\widehat{\boldsymbol{R}}_z]_{c,d}^*\} = \nonumber\\
&& \hspace{-3mm} \frac{1}{N^2} \sum_{n=0}^{N-1} \hspace{2mm}
\sum_{\underset{(0,n-N_h+1)}{r=\max}}^n
\sum_{\underset{(0,n-N_h+1)}{p=\max}}^n \hspace{1mm}
\sum_{u=0}^{N-1} \hspace{2mm}
\sum_{\underset{(0,u-N_h+1)}{s=\max}}^u \sum_{\underset{(0,u-N_h+1)}{m=\max}}^u \nonumber\\
&& \hspace{-1.5mm} h_a(n-r)h_b(n-p)h_c(u-s)h_d(u-m)\times\nonumber\\
&& \hspace{-1.5mm} \left(r_{y_a y_b}(r-p) r_{y_d y_c}(m-s) + r_{y_a
y_c}(r-s) r_{y_d y_b}(m-p)\right). \nonumber\\ \label{eq:ERzRz2}
\end{eqnarray}

The details of simplifying $\boldsymbol{U}$ for the case that the
input signal $x(t)$ is equal to the white Gaussian random process
$e(t)$ are given in Appendix \ref{sec:appndx_variance}. It is shown
that in this case, $\boldsymbol{U}$ is a diagonal matrix with
\begin{eqnarray}
\hspace{-10mm}&&[\boldsymbol{U}]_{1,1}=\frac{\sigma^4}{N^2} \Big(N^2H_1^2+(N-2N_h+2)G_1+\Sigma_1\Big)\nonumber\\
\hspace{-10mm}&&[\boldsymbol{U}]_{Q+1,Q+1}=0\label{eq:UQ+1Q+1}\nonumber\\
\hspace{-10mm}&&[\boldsymbol{U}]_{k,k}=\frac{\sigma^4}{2N^2}
\left((N-2N_h+2)G_k+\Sigma_k\right)\label{eq:Ukk3}
\end{eqnarray}
where $G_1$, $\Sigma_1$, $G_k$, and $\Sigma_k$ ($2 \leq k\leq 2Q$
and $k\neq Q+1$) are independent of the signal length and depend on
the FD filters.

The equations for computing the covariance matrix
$\mathcal{C}_{\widehat{\boldsymbol{p}}}$ as given by
\eqref{eq:cov_p_def} to \eqref{eq:Ukk3} are in the matrix form.
Next, we simplify these formulas to show the dependence of the
estimation variance on different parameters more clearly. It is
shown in Appendix \ref{sec:appndx_var_apprx} that for the white
Gaussian process, the diagonal elements of
$\mathcal{C}_{\widehat{\boldsymbol{p}}}$ can be approximated by
\begin{eqnarray}
&&\hspace{-10mm}[\mathcal{C}_{\widehat{\boldsymbol{p}}}]_{l,l}
\approx \frac{\sigma^4}{2W^2 N_x^2}\left(\frac{L^3}{Q} + L\right)
\times \nonumber \\
&& \qquad \quad \left( (N_x-2N_h L +2L)G_1 + L \Sigma_1\right)
\label{eq:var_apprx}
\end{eqnarray}
where $N_x$ is the number of Nyquist samples. Considering a large
enough $N_x$, it can be seen from \eqref{eq:var_apprx} that the
estimation variance is a cubic function of the number of spectral
segments $L$ as $(L^3/Q+L)$. Moreover, the variance is inversely
proportional to $Q$, which means that the variance decreases
quadratically with the number of sampling channels $q$. Furthermore,
at a fixed average sampling rate $(q/L)W$ and a given signal length
$N_x$, the variance increases almost linearly with the number of
spectral segments. Finally, it can be seen that the estimation
variance decreases as the signal length increases at an approximate
rate of $1/N_x$.

We next consider the asymptotic behavior of the correlogram for
undersampled data for the case of a white Gaussian process. The
following theorem studies the covariance matrix of the estimator as
the length of the signal tends to infinity. The proof of the theorem
is given in Appendix \ref{sec:appndx_thm_proof}.

\textit{Theorem 1:} In the case of a white Gaussian process, the
correlogram estimation based on undersampled data is a consistent
estimator of the average power in each spectral segment.
Furthermore, the estimations made for different spectral segments
are asymptotically uncorrelated.

\section{Numerical Examples}\label{sec:examples_sim}
In this section, we investigate the behavior of the correlogram for
undersampled data for finite-length signals based on the analytical
results obtained in Section \ref{sec:bias_var} and Monte Carlo
simulations.

The estimation bias and variance of the correlogram method depends
on the number of sampling channels $q$, the number of spectral
segments $L$, and the number of samples per channel $N$. Here, the
Nyquist sampling rate is considered to be $W=1000$ Hz. The time
offsets $c_i$ $(1\leq i \leq q)$ are distinct positive integer
numbers less than $L$ which are generated with equal probability for
each $(L,q)$-pair. After generating the time offsets $c_i$, the
matrix $\breve{\boldsymbol{\Psi}}$ is formed and its rank is
checked. In the case that $\breve{\boldsymbol{\Psi}}$ is rank
deficient, a new set of time offsets is generated until a full rank
matrix $\breve{\boldsymbol{\Psi}}$ is obtained or a maximum number
of tries is performed. In the latter case, the given $(L,q)$-pair is
considered as unfeasible. Once a full rank matrix
$\breve{\boldsymbol{\Psi}}$ is obtained, it is kept unchanged for
different signal lengths.

We present six examples to illustrate the bias and variance of the
correlogram for undersampled data. For the first four examples, we
consider a white Gaussian process with its PSD equal to $\sigma^2/W
= 1$. For the last two examples, a filtered Gaussian process is
used.

\begin{figure}[t]
\psfrag{Nx}{$N_x$} \psfrag{bias}{\scriptsize{Bias}}
\begin{center}
    \includegraphics[width=16cm]{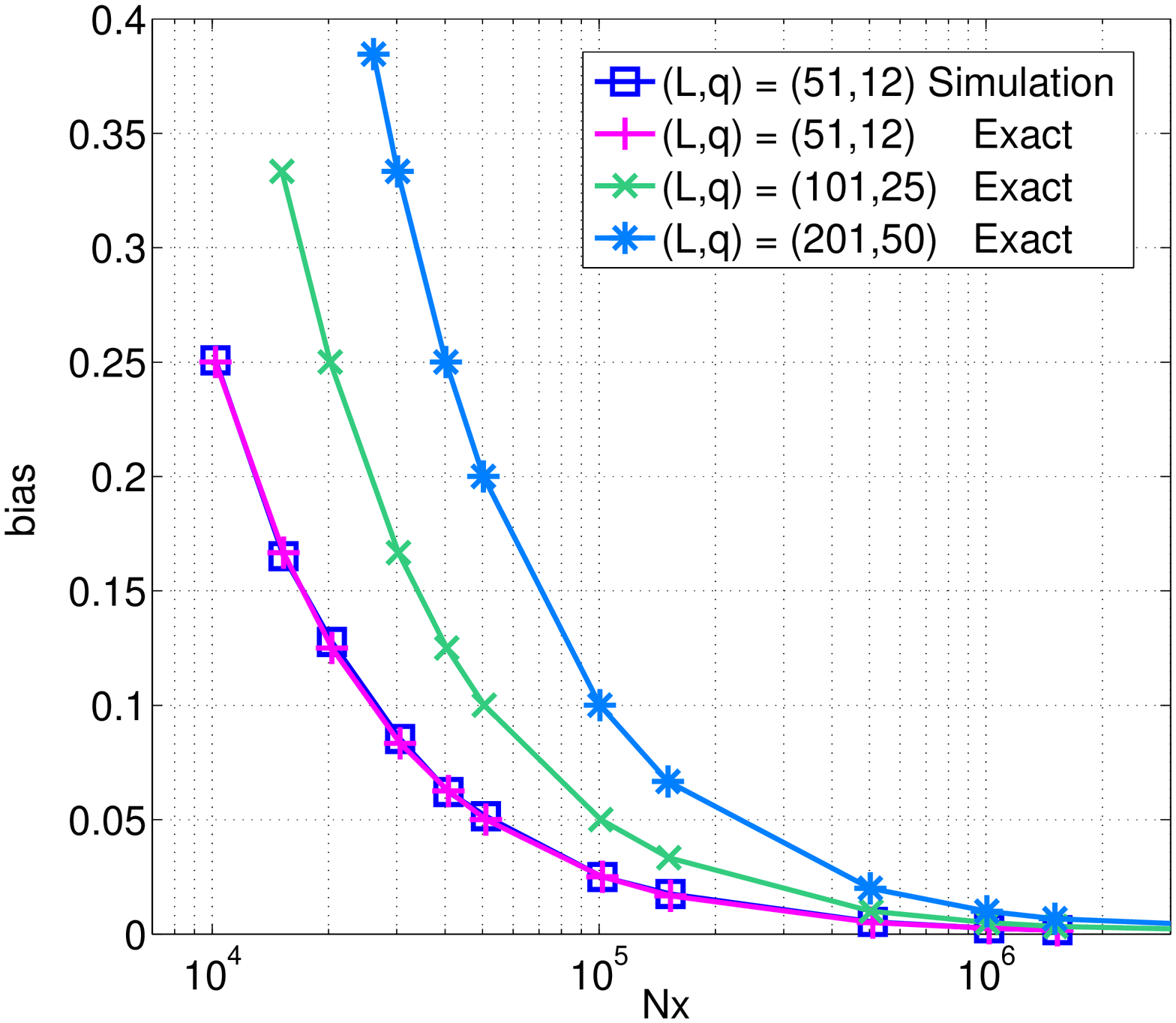}
\end{center}\vspace{-5mm}
\caption{Bias versus Nyquist signal length $N_x$. The average
sampling rate $(q/L)W$ for the $(L,q)=(51,12)$, $(101,25)$, and
$(201,50)$ pairs are $235$Hz, $247$Hz, and $248$Hz, respectively.
The curve marked with squares is obtained by Monte Carlo
simulations. The rest of the curves are based on
\eqref{eq:bias_p_white_closed}. \label{fig:bias_Nx}}\vspace{-3mm}
\end{figure}

The estimation bias is investigated first. We consider the case when
the average sampling rate $(q/L)W$ is kept unchanged. Therefore, for
a given number of Nyquist samples, the overall number of samples
available for estimation is the same for different $(L,q)$-pairs.
Fig.~\ref{fig:bias_Nx} depicts the bias of the estimator versus the
number of Nyquist samples $N_x$. The curve marked with squares is
obtained by Monte Carlo simulations for comparison with the
theoretical results. The rest of the curves are obtained from
\eqref{eq:bias_p_white_closed}. Referring to \eqref{eq:H1_def} and
\eqref{eq:bias_p_white_closed}, it can be seen that the bias is
proportional to the inverse of the signal length $N_x$ (consider
multiplying \eqref{eq:H1_def} by $L/L$, and note that $N_x=NL$).
Moreover, at a given signal length, the bias increases linearly with
the number of spectral segments. It can also be seen that the
estimation bias tends to zero as the length of the signal tends to
infinity.

\begin{figure}[t]
\psfrag{q}{$q$} \psfrag{var}{\scriptsize{Variance}}
\begin{center}
    \includegraphics[width=16cm]{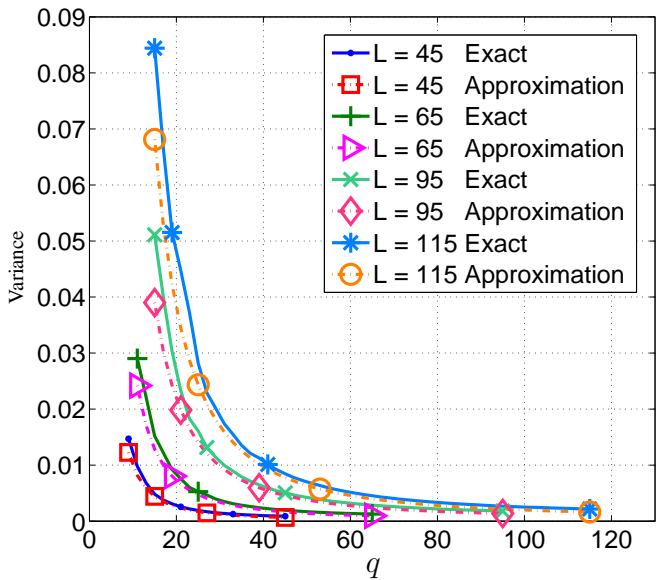}
\end{center}\vspace{-5mm}
\caption{Variance $[\mathcal{C}_{\widehat{\boldsymbol{p}}}]_{1,1}$
versus number of sampling channels $q$ at a fixed number of spectral
segments $L$. The number of Nyquist samples is set to $N_x=10^5$.
Solid lines are based on \eqref{eq:cov_p_def} to \eqref{eq:Ukk3} and
dashed lines are based on \eqref{eq:var_apprx}.
\label{fig:apprx_q}}\vspace{-3mm}
\end{figure}

Fig.~\ref{fig:apprx_q} depicts the variance of the estimator
$[\mathcal{C}_{\widehat{\boldsymbol{p}}}]_{1,1}$ versus the number
of sampling channels $q$ for different values of spectral segments
$L$. The signal length is fixed at $N_x=10^5$. The curves drawn with
solid lines represent the exact variance obtained from
\eqref{eq:cov_p_def} to \eqref{eq:Ukk3}, and the curves plotted with
dashed lines are the approximate values obtained from
\eqref{eq:var_apprx}. Increasing $q$ at a fixed $L$ is equivalent to
increasing the average sampling rate $(q/L)W$. According to the
approximate variance as given in \eqref{eq:var_apprx}, the variance
decreases quadratically with the number of sampling channels $q$.
Therefore, the performance of the estimator improves by increasing
$q$, but this comes at the price of adding to the complexity of the
system by using more sampling channels.

\begin{figure}[t]
\psfrag{L}{$L$}\psfrag{var}{\scriptsize{Variance}}
\begin{center}
    \includegraphics[width=16cm]{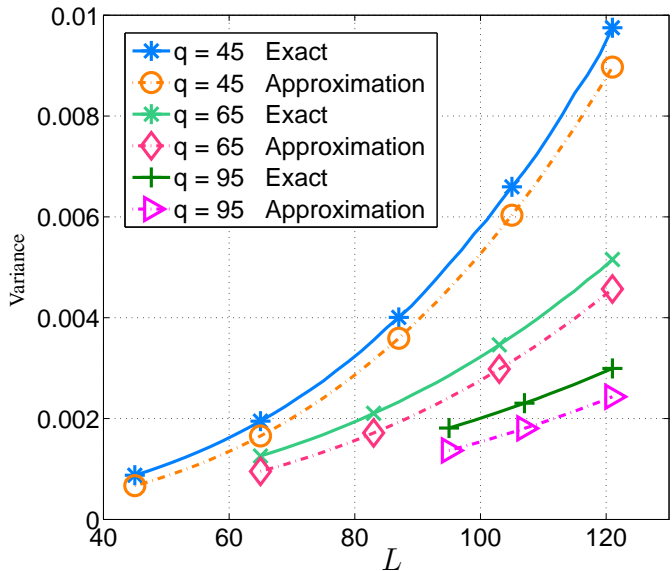}
\end{center}\vspace{-5mm}
\caption{Variance $[\mathcal{C}_{\widehat{\boldsymbol{p}}}]_{1,1}$
versus number of spectral segments $L$ at a fixed number of sampling
channels $q$. The number of Nyquist samples is set to $N_x=10^5$.
Solid lines are based on \eqref{eq:cov_p_def} to \eqref{eq:Ukk3} and
dashed lines are based on \eqref{eq:var_apprx}.
\label{fig:apprx_L}}\vspace{-3mm}
\end{figure}

Fig.~\ref{fig:apprx_L} shows the variance of the estimator
$[\mathcal{C}_{\widehat{\boldsymbol{p}}}]_{1,1}$ versus the number
of spectral segments $L$ for different numbers of  sampling channels
$q$. The signal length is fixed at $N_x=10^5$. Again, the curves
drawn with solid lines are obtained from \eqref{eq:cov_p_def} to
\eqref{eq:Ukk3}, and the curves plotted with dashed lines are
obtained from \eqref{eq:var_apprx}. According to the approximate
variance as given in \eqref{eq:var_apprx}, the variance increases
cubicly with the number of spectral segments $L$. Therefore, at a
fixed signal length and fixed number of sampling channels, the
performance of the estimator is degraded by increasing the number of
spectral segments $L$, i.e., by increasing the frequency resolution.

\begin{figure}[t]
\psfrag{Nx}{$N_x$} \psfrag{var}{\scriptsize{Variance}}
\begin{center}
    \includegraphics[width=16cm]{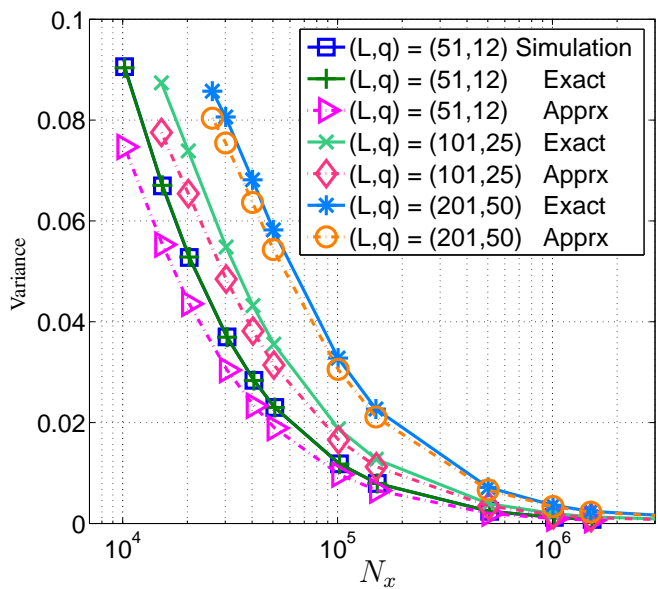}
\end{center}\vspace{-5mm}
\caption{Variance $[\mathcal{C}_{\widehat{\boldsymbol{p}}}]_{1,1}$
versus Nyquist signal length $N_x$. The average sampling rate
$(q/L)W$ for the $(L,q)=(51,12)$, $(101,25)$, and $(201,50)$ pairs
are $235$Hz, $247$Hz, and $248$Hz, respectively. The curve marked
with squares is obtained by Monte Carlo simulations. Solid lines are
based on \eqref{eq:cov_p_def} to \eqref{eq:Ukk3}, and dashed lines
are based on \eqref{eq:var_apprx}. \label{fig:var_Nx}}\vspace{-3mm}
\end{figure}

The variance of the estimator
$[\mathcal{C}_{\widehat{\boldsymbol{p}}}]_{1,1}$ versus the signal
length $N_x$ is illustrated in Fig.~\ref{fig:var_Nx}. Here, the
average sampling rate $(q/L)W$ is kept unchanged. Therefore, for a
given number of Nyquist samples, the overall number of samples
available for estimation is the same for different $(L,q)$-pairs.
The curve marked with squares is obtained by Monte Carlo simulations
for comparison with the theoretical results. Again, the curves drawn
with solid lines are obtained from \eqref{eq:cov_p_def} to
\eqref{eq:Ukk3}, and the curves plotted with dashed lines are
obtained from \eqref{eq:var_apprx}. Referring to the approximate
variance as given in \eqref{eq:var_apprx}, the variance is almost
proportional to the inverse of the signal length $N_x$. From the
curves corresponding to the $(51,12)$, $(101,25)$, and
$(201,50)$-pairs in Fig.~\ref{fig:var_Nx}, it can be seen that the
performance of the estimator degrades when increasing the number of
spectral segments, i.e., when increasing the frequency resolution.
The average sampling rate is kept almost the same in this scenario.
It can also be seen that the estimation variance tends to zero as
the length of the signal tends to infinity.

For the next two examples, we consider a more general case with a
filtered Gaussian process. The signal is obtained by passing a white
Gaussian signal through a bandlimited filter with cutoff frequencies
set at $W/10$ and $W/5$ Hz. Through our experiments, we found that
the estimation variance at each spectral segment depends not only on
the power of signal at that frequency band, but also it is dependant
on the power of the signal at other spectral segments. As noticed
from the analytical derivations for the white Gaussian process (see
\eqref{eq:EppT}, \eqref{eq:Ukk3}, and \eqref{eq:var_apprx}), the
estimation variance is proportional to the square of the signal
power ($\sigma^4/W^2$). Therefore, we set the gain of the filter so
that the square of the power averaged over all spectral segments for
both the white Gaussian process at the input of the filter and the
filtered signal is the same.

\begin{figure}[t]
\psfrag{L}{$L$} \psfrag{var}{\scriptsize{Variance}}
\begin{center}
    \includegraphics[width=16cm]{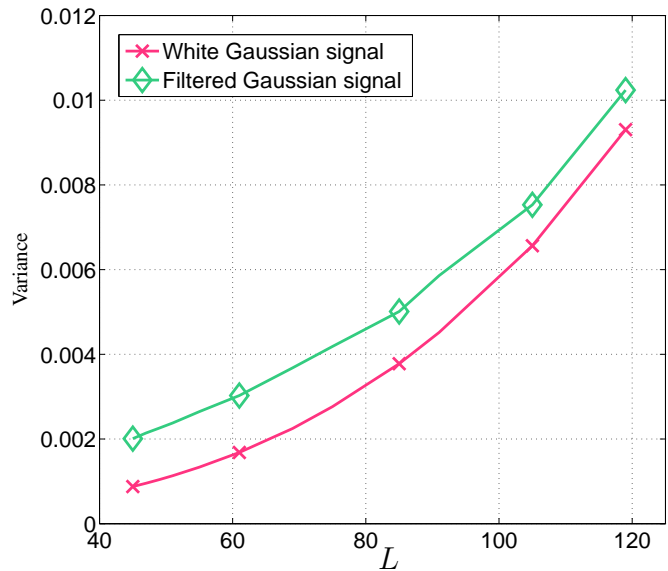}
\end{center}\vspace{-5mm}
\caption{Variance $[\mathcal{C}_{\widehat{\boldsymbol{p}}}]_{1,1}$
versus number of spectral segments $L$ at a fixed number of sampling
channels $q=45$. The number of Nyquist samples is set to $N_x=10^5$.
The curve for the white Gaussian signal is based on
\eqref{eq:cov_p_def} to \eqref{eq:Ukk3}, and the curve for the
filtered Gaussian signal is obtained by Monte Carlo simulations.
\label{fig:general_L}}\vspace{-3mm}
\end{figure}

In Fig.~\ref{fig:general_L}, the variance of the estimator
$[\mathcal{C}_{\widehat{\boldsymbol{p}}}]_{1,1}$ versus the number
of spectral segments $L$ is depicted. The number of sampling
channels is set to $q=45$, and the signal length is fixed at
$N_x=10^5$. The curve for the white Gaussian signal is based on
\eqref{eq:cov_p_def} to \eqref{eq:Ukk3}, and the curve for the
filtered Gaussian signal is obtained by Monte Carlo simulations. The
latter curve is the average estimation variance of the spectral
segments that pass through the filter. It can be seen in
Fig.~\ref{fig:general_L} that the variance of the estimator for the
white and the filtered signals are close to each other.

\begin{figure}[t]
\psfrag{Nx}{$N_x$} \psfrag{var}{\scriptsize{Variance}}
\begin{center}
    \includegraphics[width=16cm]{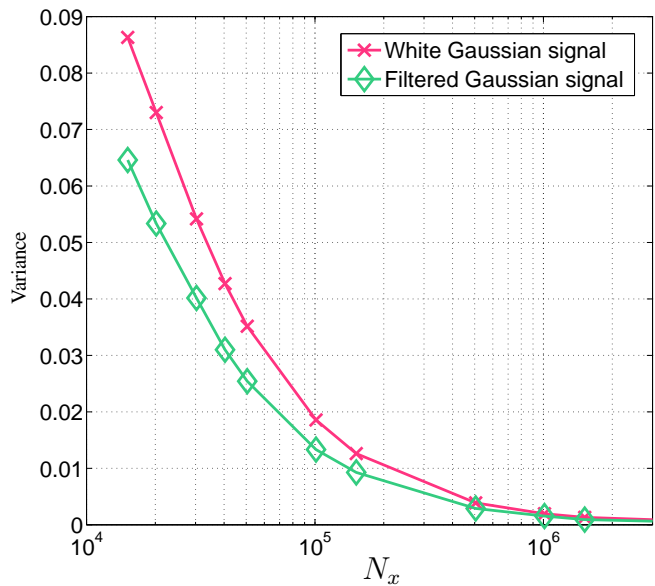}
\end{center}\vspace{-5mm}
\caption{Variance $[\mathcal{C}_{\widehat{\boldsymbol{p}}}]_{1,1}$
versus Nyquist signal length $N_x$ for $(L,q)=(101,25)$ pair. The
curve for the white Gaussian signal is based on \eqref{eq:cov_p_def}
to \eqref{eq:Ukk3}, and the curve for the filtered Gaussian signal
is obtained by Monte Carlo simulations.
\label{fig:general_var_Nx}}\vspace{-3mm}
\end{figure}

Finally, the variance of the estimator
$[\mathcal{C}_{\widehat{\boldsymbol{p}}}]_{1,1}$ versus the signal
length $N_x$ for the white and the filtered signals is investigated.
The number of spectral segments is set to $L=101$, and the number of
sampling channels is set to $q=25$. Again, the curve for the white
Gaussian signal is based on \eqref{eq:cov_p_def} to \eqref{eq:Ukk3},
and the curve for the filtered Gaussian signal is obtained by Monte
Carlo simulations. Similar to the previous example, it can be seen
in Fig.~\ref{fig:general_var_Nx} that the estimation variance for
the white and the filtered signals are close to each other. It can
also be seen that the estimation variance tends to zero as the
length of the signal tends to infinity.

\section{Conclusion}\label{sec:conclude}
We considered the correlogram for undersampled data which estimates
the spectrum from a subset of the Nyquist samples. This method has
been analyzed in this paper by computing the bias and the variance
of the estimator. It has been shown that the bias and the variance
of the method tend to zero asymptotically. Therefore, this method is
a consistent estimator. Furthermore, it has been shown that the
estimation made for different spectral segments becomes uncorrelated
as the signal length goes to infinity.

The behavior of the estimator for finite-length signals has also
been investigated. It has been shown that at a given signal length,
the estimation accuracy increases as the average sampling rate is
increased (either by decreasing the frequency resolution $L$ or by
increasing the complexity of the system $q$). It has also been shown
that at a fixed average sampling rate, the performance of the
estimator degrades for the estimation with higher frequency
resolution. To sum up, it has been illustrated that there is a
tradeoff between the accuracy of the estimator (the estimation
variance), the frequency resolution (the number of spectral
segments), and the complexity of the estimator (the number of
sampling channels).

\appendices
\section{Bias Simplification}\label{sec:appndx_bias}
In the case that $x(t)$ is equal to $e(t)$, we have
$h_x(m)=\delta(m)$. Then, using \eqref{eq:r_yayb_e}, the
cross-correlation function $r_{y_a y_b}(k)$ is given by
\begin{equation}
r_{y_a y_b}(k) = \sigma^2 \delta(k) \delta(a-b).\label{eq:r_yayb_e2}
\end{equation}
Applying \eqref{eq:r_yayb_e2} to \eqref{eq:ERz01}, we find that
\begin{equation}
E\{[\widehat{\boldsymbol{R}}_{\boldsymbol{z}}]_{a,b}\} =
[\boldsymbol{R}_{\boldsymbol{z}}]_{a,b} - \frac{1}{N}
\sum_{r=0}^{N_h-1} h_a^2(r) r \sigma^2 \delta(a-b). \label{eq:ERz_e}
\end{equation}
Next, $[\boldsymbol{R}_{\boldsymbol{z}}]_{a,b}$ is obtained using
\eqref{eq:Rz_ab_D} and \eqref{eq:r_yayb_e2} as
\begin{equation}
[\boldsymbol{R}_{\boldsymbol{z}}]_{a,b} = \sum_{r=0}^{N_h-1}
h_a^2(r) \sigma^2 \delta(a-b). \label{eq:Rz_ab_e}
\end{equation}
Replacing \eqref{eq:Rz_ab_e} into \eqref{eq:ERz_e} results in
\begin{equation}
E\{[\widehat{\boldsymbol{R}}_z]_{a,b}\} = 0 \label{eq:ERz2}
\end{equation}
for $a\neq b$, and
\begin{equation}
E\{[\widehat{\boldsymbol{R}}_z]_{a,b}\} = \sigma^2 \frac{1}{N}
\sum_{r=0}^{N_h-1} (N - r) h_a^2(r) = H_a \sigma^2 \label{eq:ERz3}
\end{equation}
for $a = b$, where
\begin{equation}
H_a \triangleq \frac{1}{N}\sum_{r=0}^{N_h-1}(N-r)h_a^2(r).
\label{eq:Ha_def}
\end{equation}
Recalling that the first diagonal element of
$\widehat{\boldsymbol{R}}_z$ is used in
$\widehat{\breve{\boldsymbol{u}}}$ and taking the real and imaginary
parts of \eqref{eq:ERz2} and \eqref{eq:ERz3},
$E\{\widehat{\breve{\boldsymbol{u}}}\}$ can be obtained as
\begin{equation}
E\{\widehat{\breve{\boldsymbol{u}}}\} =
H_1\sigma^2\boldsymbol{e}_1\label{eq:Eu_brev}
\end{equation}
where $\boldsymbol{e}_1$ is a column vector of length $q(q-1)+2$
with all its elements equal to zero except for the first element
which is $1$. The expected value of $\widehat{\boldsymbol{v}}$ can
be found using \eqref{eq:Ev1} and \eqref{eq:Eu_brev} as
\begin{equation}
E\{\widehat{\boldsymbol{v}}\} =
H_1\sigma^2(\breve{\boldsymbol{\Psi}}^T\breve{\boldsymbol{\Psi}})^{-1}\breve{\boldsymbol{\Psi}}^T\boldsymbol{e}_1.\label{eq:Evhat}
\end{equation}

Next, Consider the fact that $x(t)$ has equal power in all spectral
segments (the elements of $\boldsymbol{v}$ are all the same). Since
$\widehat{\boldsymbol{v}}$ is asymptotically unbiased, it follows
that the elements of
$\lim_{N\rightarrow\infty}E\{\widehat{\boldsymbol{v}}\}$ are also
equal.

Replacing the true values in \eqref{eq:base_cov_elmnt} with the
estimated values for $a=b=1$, taking expectation from both sides,
and letting the number of samples tend to infinity, we obtain that
\begin{eqnarray}
\lim_{N\rightarrow\infty}
E\{[\widehat{\boldsymbol{R}}_z]_{1,1}\}\hspace{-2mm} &=& \hspace{-2mm} \sum_{l=1}^L\lim_{N\rightarrow\infty}E\{\widehat{v}_l\}\nonumber\\
\hspace{-2mm}&=&\hspace{-2mm}
\boldsymbol{1}_L^T\lim_{N\rightarrow\infty}E\{\widehat{\boldsymbol{v}}\}\label{eq:limRz1}
\end{eqnarray}
where $\widehat{v}_l$ ($1\leq l\leq L$) are the elements of
$\widehat{\boldsymbol{v}}$. Considering normalized FD filters
($\sum_{r=0}^{N_h-1}h_a^2(r)=1$) and referring to \eqref{eq:Ha_def},
we also find that
\begin{equation}
\lim_{N\rightarrow\infty}H_a=1.\label{eq:limHa}
\end{equation}
Therefore, using \eqref{eq:ERz3}, we can find that
\begin{equation}
\lim_{N\rightarrow\infty}E\{[\widehat{\boldsymbol{R}}_z]_{1,1}\} =
\sigma^2.\label{eq:limRz2}
\end{equation}
Combining \eqref{eq:limRz1} with \eqref{eq:limRz2} results in
\begin{equation}
\lim_{N\rightarrow\infty}E\{\widehat{\boldsymbol{v}}\}=\frac{\sigma^2}{L}\boldsymbol{1}_L.\label{eq:limEvhat}
\end{equation}
Letting the number of samples tend to infinity in \eqref{eq:Evhat}
and using \eqref{eq:limEvhat}, we obtain
\begin{equation}
\lim_{N\rightarrow\infty}E\{\widehat{\boldsymbol{v}}\}=
\sigma^2(\breve{\boldsymbol{\Psi}}^T\breve{\boldsymbol{\Psi}})^{-1}\breve{\boldsymbol{\Psi}}^T\boldsymbol{e}_1
= \frac{\sigma^2}{L} \boldsymbol{1}_L.\label{eq:limEvhat2}
\end{equation}
It follows from \eqref{eq:limEvhat2} that all the elements of the
first column of
$(\breve{\boldsymbol{\Psi}}^T\breve{\boldsymbol{\Psi}})^{-1}\breve{\boldsymbol{\Psi}}^T$
are equal to $1/L$. Therefore, \eqref{eq:Evhat} can be simplified as
\begin{equation}
E\{\widehat{\boldsymbol{v}}\}=H_1\frac{\sigma^2}{L}\boldsymbol{1}_L.
\end{equation}
Finally, using \eqref{eq:p_hat_def}, we have
\begin{equation}
E\{\widehat{\boldsymbol{p}}\} =
H_1\frac{\sigma^2}{W}\boldsymbol{1}_L.
\end{equation}

\section{Variance Simplification}\label{sec:appndx_variance}
In the case that $x(t)$ is equal to $e(t)$, we have
$h_x(m)=\delta(m)$. Then, the cross-correlation functions in
\eqref{eq:ERzRz1} are simplified as
\begin{eqnarray}
E_1\hspace{-2mm}&\triangleq&\hspace{-2mm} r_{y_a y_b}(r-p) r_{y_c
y_d}(s-m) + r_{y_a y_d}(r-m) r_{y_c y_b}(s-p) \nonumber\\
\hspace{-2mm}&=&\hspace{-2mm}\sigma^4\big(\delta(r-p)\delta(a-b)\delta(s-m)\delta(c-d)+\nonumber\\
&&~\quad
\delta(r-m)\delta(a-d)\delta(s-p)\delta(c-b)\big).\label{eq:E1def}
\end{eqnarray}
Similarly, the cross-correlation functions in \eqref{eq:ERzRz2} are
simplified as
\begin{eqnarray}
E_2\hspace{-2mm} &\triangleq& \hspace{-2mm}r_{y_a y_b}(r-p) r_{y_d
y_c}(m-s) + r_{y_a y_c}(r-s) r_{y_d y_b}(m-p)\nonumber\\
\hspace{-2mm} &=& \hspace{-2mm} \sigma^4\big(\delta(r-p)\delta(a-b)\delta(m-s)\delta(d-c)+\nonumber\\
\hspace{-2mm}&&~ \quad
\delta(r-s)\delta(a-c)\delta(m-p)\delta(d-b)\big).
\end{eqnarray}

Recalling that only the first diagonal element of
$\widehat{\boldsymbol{R}}_z$ is present in
$\widehat{\breve{\boldsymbol{u}}}$, $E_1$ can be found to be equal
to
\begin{equation}
E_1=\sigma^4\big(\delta(r-p)\delta(s-m)+\delta(r-m)\delta(s-p)\big)\label{eq:E1}
\end{equation}
for $a=b=c=d=1$, and it equals to zero otherwise. Similarly, $E_2$
can be found to be equal to
\begin{equation}
E_2=\sigma^4\delta(r-s)\delta(m-p)\label{eq:E2}
\end{equation}
for $a=c$ and $b=d$, and it equals zero otherwise (excluding the
case when $a=b=c=d=1$ since $[\widehat{\boldsymbol{R}}_z]_{1,1}$ is
real-valued, and therefore, we do not need to compute
\eqref{eq:ERzRz2}). Noting that
$E\{[\widehat{\boldsymbol{R}}_z]_{a,b}[\widehat{\boldsymbol{R}}_z]_{c,d}\}$
and
$E\{[\widehat{\boldsymbol{R}}_z]_{a,b}[\widehat{\boldsymbol{R}}_z]_{c,d}^*\}$
are real-valued and using \eqref{eq:RexImy}, \eqref{eq:E1}, and
\eqref{eq:E2}, we can find that all the off-diagonal elements of
$\boldsymbol{U}$ are equal to zero.

Let us start computing the diagonal elements of $\boldsymbol{U}$ by
setting $a=b=c=d=1$. It follows from \eqref{eq:ERzRz1} and
\eqref{eq:E1} that
\begin{eqnarray}
&& \hspace{-7mm}
E\{[\widehat{\boldsymbol{R}}_z]_{1,1}[\widehat{\boldsymbol{R}}_z]_{1,1}\}
= \frac{\sigma^4}{N^2} \Big( \sum_{n=0}^{N-1} \hspace{1mm}
\sum_{\underset{(0,n-N_h+1)}{r=\max}}^n \hspace{1mm}
\sum_{u=0}^{N-1} \hspace{1mm}
\sum_{\underset{(0,u-N_h+1)}{s=\max}}^u \nonumber\\
&& \quad \quad \qquad \qquad h_1^2(n-r) h_1^2(u-s) +
\sum_{n=0}^{N-1} S_1(n)\Big) \label{eq:ERzRz3}
\end{eqnarray}
where $S_1(n)$ is defined as
\begin{eqnarray}
&& \hspace{-5mm} S_1(n) \triangleq \hspace{-2mm}
\sum_{\underset{(0,n-N_h+1)}{r=\max}}^n
\sum_{\underset{(0,n-N_h+1)}{p=\max}}^n \hspace{1mm}
\sum_{u=0}^{N-1} \hspace{1mm}
\sum_{\underset{(0,u-N_h+1)}{s=\max}}^u
\sum_{\underset{(0,u-N_h+1)}{m=\max}}^u \nonumber\\
&& \hspace{-3mm} \delta(r-m)\delta(s-p)
h_1(n-r)h_1(n-p)h_1(u-s)h_1(u-m). \nonumber\\
\end{eqnarray}
For $N_h-1\leq n\leq N-N_h$, $S_1(n)$ is given by
\begin{eqnarray}
S_1(n) \hspace{-2mm} &=& \hspace{-2mm}
\sum_{\underset{n-N_h+1}{u=}}^{n+N_h-1}
\Bigg[\sum_{\underset{n-N_h+1}{r=}}^n
\sum_{\underset{(0,u-N_h+1)}{m=\max}}^u \delta(r-m) \times \nonumber\\
&& \hspace{30mm}
h_1(n-r) h_1(u-m) \Bigg] \times \nonumber\\
&& \hspace{10mm} \Bigg[ \sum_{\underset{n-N_h+1}{p=}}^n \hspace{1mm}
\sum_{\underset{(0,u-N_h+1)}{s=\max}}^u \delta(s-p) \times \nonumber\\
&& \hspace{30mm} h_1(n-p) h_1(u-s) \Bigg].
\end{eqnarray}
Note that the summations in the brackets are equivalent to each
other, which leads to the following simplification
\begin{eqnarray}
S_1(n) \hspace{-2mm} &=& \hspace{-2mm}
\sum_{\underset{n-N_h+1}{u=}}^{n+N_h-1}
\Bigg[\sum_{\underset{n-N_h+1}{r=}}^n
\sum_{\underset{(0,u-N_h+1)}{m=\max}}^u \delta(r-m) \times \nonumber\\
&& \hspace{30mm} h_1(n-r) h_1(u-m) \Bigg]^2 \nonumber\\
&=& \sum_{\underset{n-N_h+1}{u=}}^{n+N_h-1}
\Bigg[\sum_{r=\max(n,u)-N_h+1}^{\min(n,u)} \nonumber\\
&& \hspace{27mm} h_1(n-r) h_1(u-r) \Bigg]^2.
\end{eqnarray}
Next, a change of variable ($g=u-n+N_h-1$) is used, which results in
\begin{eqnarray}
S_1(n) \hspace{-2mm} &=& \hspace{-2mm} \sum_{g=0}^{2N_h-2}
\Bigg[\sum_{r=\max(0,g-N_h+1)+n-N_h+1}^{\min(0,g-N_h+1)+n} \nonumber\\
&& \hspace{2mm} h_1(n-r) h_1(n-r+g-N_h+1) \Bigg]^2.
\end{eqnarray}
With another change of variable ($p=n-r+g-N_h+1$), we obtain the
following
\begin{eqnarray}
S_1(n) \hspace{-2mm} &=& \hspace{-2mm} \sum_{g=0}^{2N_h-2}
\Bigg[\sum_{p=\max(0,g-N_h+1)}^{\min(g,N_h-1)} \nonumber\\
&& \hspace{20mm} h_1(p-g+N_h-1) h_1(p) \Bigg]^2
\end{eqnarray}
which is equal to
\begin{equation}
G_1\triangleq
S_1(n)=\sum_{g=0}^{2N_h-2}\left[h_1(i)*h_1(N_h-1-i)\vert_g\right]^2
\end{equation}
where $*$ denotes the convolution operation. Note that $G_1$ is not
a function of $n$. In a similar way, $S_1(n)$ for $0\leq n<N_h-1$ is
given by
\begin{equation}
S_1(n)=\sum_{g=0}^{n+N_h-1}\left[(h_1(i)W_n(i))*h_1(N_h-1-i)\vert_g\right]^2
\end{equation}
where $W_n(i)$ is equal to $1$ for $0\leq i \leq n$ and zero
elsewhere. For $N-N_h< n \leq N-1$, $S_1(n)$ is given by
\begin{equation}
S_1(n)=\sum_{g=0}^{N-n+N_h-2}\left[h_1(i)*h_1(N_h-1-i)\vert_g\right]^2.
\end{equation}
Next, \eqref{eq:ERzRz3} can be rewritten as
\begin{eqnarray}
&&\hspace{-8mm}E\{[\widehat{\boldsymbol{R}}_z]_{1,1}[\widehat{\boldsymbol{R}}_z]_{1,1}\}
= \frac{\sigma^4}{N^2} \times\nonumber\\
&& \hspace{-5mm} \Big(\sum_{n=0}^{N-1}
\sum_{\underset{(0,n-N_h+1)}{r=\max}}^n \hspace{-2mm} h_1^2(n-r)
\sum_{u=0}^{N-1} \hspace{1mm}
\sum_{\underset{(0,u-N_h+1)}{s=\max}}^u h_1^2(u-s)+\nonumber\\
&& \hspace{-4mm}
(N-2N_h+2)G_1+\sum_{n=0}^{N_h-2}S_1(n)+\hspace{-5mm}\sum_{n=N-N_h+1}^{N-1}S_1(n)\Big).
\label{eq:ERzRz5}
\end{eqnarray}
Using \eqref{eq:Ha_def}, we have
\begin{equation}
\sum_{n=0}^{N-1} \sum_{\underset{(0,n-N_h+1)}{r=\max}}^n
\hspace{-2mm} h_1^2(n-r) = \sum_{r=0}^{N_h-1}(N-r)h_1^2(r) = N H_1.
\end{equation}
Therefore, \eqref{eq:ERzRz5} can be simplified as
\begin{equation}
E\{[\widehat{\boldsymbol{R}}_z]_{1,1}[\widehat{\boldsymbol{R}}_z]_{1,1}\}
= \frac{\sigma^4}{N^2}
\Big(N^2H_1^2+(N-2N_h+2)G_1+\Sigma_1\Big)\label{eq:ERzRz4}
\end{equation}
where $\Sigma_1
\triangleq\sum_{n=0}^{N_h-2}S_1(n)+\sum_{n=N-N_h+1}^{N-1}S_1(n)$.
Note that $[\widehat{\boldsymbol{R}}_z]_{1,1}$ is real-valued.
Therefore, $[\boldsymbol{U}]_{1,1}$ is equal to
$E\{[\widehat{\boldsymbol{R}}_z]_{1,1}[\widehat{\boldsymbol{R}}_z]_{1,1}\}$
as given in \eqref{eq:ERzRz4} and $[\boldsymbol{U}]_{Q+1,Q+1}$
equals zero since the imaginary part of
$[\widehat{\boldsymbol{R}}_z]_{1,1}$ is zero.

For the rest of the diagonal elements of $\boldsymbol{U}$,
$E\{[\widehat{\boldsymbol{R}}_z]_{a,b}[\widehat{\boldsymbol{R}}_z]_{a,b}\}$
equals zero, as $E_1$ is zero. Therefore, $[\boldsymbol{U}]_{k,k}$
($2 \leq k\leq 2Q$ and $k\neq Q+1$) can be obtained using
\eqref{eq:RexRey} and \eqref{eq:ImxImy} as
\begin{equation}
[\boldsymbol{U}]_{k,k}=\frac{1}{2}Re\left(E\{[\widehat{\boldsymbol{R}}_z]_{a,b}[\widehat{\boldsymbol{R}}_z]_{a,b}^*\}\right).
\end{equation}
From \eqref{eq:ERzRz2} and \eqref{eq:E2} we have
\begin{equation}
[\boldsymbol{U}]_{k,k}=\frac{\sigma^4}{2N^2}
\sum_nS_k(n)\label{eq:Ukk}
\end{equation}
where $S_k(n)$ is defined as
\begin{eqnarray}
&& \hspace{-8mm} S_k(n)\triangleq
\sum_{\underset{(0,n-N_h+1)}{r=\max}}^n
\sum_{\underset{(0,n-N_h+1)}{p=\max}}^n \hspace{1mm}
\sum_{u=0}^{N-1} \hspace{2mm}
\sum_{\underset{(0,u-N_h+1)}{s=\max}}^u
\sum_{\underset{(0,u-N_h+1)}{m=\max}}^u \nonumber\\
&& \hspace{-6mm} \delta(r-s)\delta(m-p)
h_a(n-r)h_b(n-p)h_a(u-s)h_b(u-m). \nonumber\\
\end{eqnarray}
It can be shown that for $N_h-1\leq n\leq N-N_h$, $S_k(n)$ is given
by
\begin{eqnarray}
&& \hspace{-8mm} G_k\triangleq
S_k(n)=\sum_{g=0}^{2N_h-2}\left(h_a(i)*h_a(N_h-1-i)\right)\vert_g\times\nonumber\\
&&\qquad \qquad \qquad \left(h_b(i)*h_b(N_h-1-i)\right)\vert_g.
\end{eqnarray}
For $0\leq n<N_h-1$, $S_k(n)$ is given by
\begin{eqnarray}
\hspace{-13mm}&&S_k(n)=\sum_{g=0}^{n+N_h-1}\left((h_a(i)W_n(i))*h_a(N_h-1-i)\right)\vert_g\times\nonumber\\
\hspace{-13mm}&&\qquad \qquad \qquad \quad
\left((h_b(i)W_n(i))*h_b(N_h-1-i)\right)\vert_g.
\end{eqnarray}
For $N-N_h< n \leq N-1$, $S_k(n)$ is given by
\begin{eqnarray}
&& \hspace{-5mm} S_k(n)=\sum_{g=0}^{N-n+N_h-2}\left(h_a(i)*h_a(N_h-1-i)\right)\vert_g\times\nonumber\\
&& \qquad \qquad \qquad \left(h_b(i)*h_b(N_h-1-i)\right)\vert_g.
\end{eqnarray}
Thus, \eqref{eq:Ukk} can be rewritten as
\begin{equation}
[\boldsymbol{U}]_{k,k}=\frac{\sigma^4}{2N^2}
\left((N-2N_h+2)G_k+\Sigma_k\right)\label{eq:Ukk2}
\end{equation}
where
$\Sigma_k\triangleq\sum_{n=0}^{N_h-2}S_k(n)+\sum_{n=N-N_h+1}^{N-1}S_k(n)$.

\section{Variance Approximation}\label{sec:appndx_var_apprx}
Referring to \eqref{eq:EppT}, computation of the $l$-th diagonal
element of the covariance matrix requires the knowledge of the
elements of the $l$-th row of $\boldsymbol{A} \triangleq
(\breve{\boldsymbol{\Psi}}^T\breve{\boldsymbol{\Psi}})^{-1}\breve{\boldsymbol{\Psi}}^T$.
The diagonal elements of $\boldsymbol{U}$ for $2 \leq k \leq 2Q$ and
$k \neq Q+1$ as given by \eqref{eq:Ukk3} differ from each other in
$G_k$ and $\Sigma_k$. However, the values of $G_k$ and $\Sigma_k$
for different values of $k$ almost remain the same as they are
related to the energy of the FD filters which are normalized to one.
Let us approximate $G_k$ and $\Sigma_k$ by $G_1$ and $\Sigma_1$.
Then, $[\boldsymbol{U}]_{k,k}$ can be approximated by
\begin{equation}
\gamma \triangleq \frac{\sigma^4}{2N^2}
\left((N-2N_h+2)G_1+\Sigma_1\right).\label{eq:Ukk4}
\end{equation}
The approximation in \eqref{eq:Ukk4} relaxes the problem of
computing the $l$-th diagonal element of the covariance matrix to
just finding the Euclidean norm of the $l$-th row of
$\boldsymbol{A}$. The squared norm of the $l$-th row of
$\boldsymbol{A}$ can be obtained as
\begin{eqnarray}
\phi_l &\triangleq& \left[\boldsymbol{A} \boldsymbol{A}^T\right]_{l,l} \nonumber \\
&=&
\left[(\breve{\boldsymbol{\Psi}}^T\breve{\boldsymbol{\Psi}})^{-1}\right]_{l,l} \nonumber \\
&=&
\left[\left(Re(\boldsymbol{\Psi}^H\boldsymbol{\Psi})\right)^{-1}\right]_{l,l}.
\end{eqnarray}
Referring to \eqref{eq:Psi_elements}, the diagonal elements of
$Re(\boldsymbol{\Psi}^H\boldsymbol{\Psi})$ are all equal to $Q$, and
the off-diagonal elements are given as
\begin{equation}
\left[Re(\boldsymbol{\Psi}^H\boldsymbol{\Psi})\right]_{i,j}=1+\sum_{k=2}^Q
\cos((i-j)\omega_k)
\end{equation}
where $1 \leq j, j \leq L$ and $i \neq j$. Noting that the
frequencies $\omega_k$ are randomly obtained based on the sampling
pattern, the value of the off-diagonal elements of
$Re(\boldsymbol{\Psi}^H\boldsymbol{\Psi})$ are negligible compared
to the value of the diagonal elements. Therefore,
$Re(\boldsymbol{\Psi}^H\boldsymbol{\Psi})$ can be approximated by a
diagonal matrix with elements equal to $Q$, which results in
\begin{equation}
\phi_l \approx \frac{1}{Q}.\label{eq:phi_apprx}
\end{equation}

It is shown in Appendix \ref{sec:appndx_bias} that all the elements
of the first column of $\boldsymbol{A}$ are equal to $1/L$.
Furthermore, all the elements of the $(Q+1)$-th column of
$\boldsymbol{A}$ are equal to zero, as all the elements of the
$(Q+1)$-th row of $\breve{\boldsymbol{\Psi}}$ are zero. Then, using
\eqref{eq:Ephat}, \eqref{eq:cov_p_def}, \eqref{eq:EppT}, and
\eqref{eq:Ukk4}, $[\mathcal{C}_{\widehat{\boldsymbol{p}}}]_{l,l}$
can be approximated as
\begin{equation}
[\mathcal{C}_{\widehat{\boldsymbol{p}}}]_{l,l} \approx
\left(\frac{L}{W}\right)^2 \bigg[ \gamma \phi_l +
\frac{1}{L^2}\left([\boldsymbol{U}]_{1,1}-\gamma\right) \bigg] -
\left(H_1\frac{\sigma^2}{W}\right)^2. \label{eq:cov_p_apprx}
\end{equation}
Next, using \eqref{eq:Ukk3}, \eqref{eq:Ukk4}, and
\eqref{eq:phi_apprx}, we can simplify \eqref{eq:cov_p_apprx} to
\begin{eqnarray}
&&\hspace{-10mm}[\mathcal{C}_{\widehat{\boldsymbol{p}}}]_{l,l}
\approx \frac{\sigma^4}{2W^2 N_x^2}\left(\frac{L^3}{Q} + L\right)
\times \nonumber \\
&& \qquad \quad \left( (N_x-2N_h L +2L)G_1 + L \Sigma_1\right)
\end{eqnarray}
where $N_x \triangleq NL $ is the number of Nyquist samples.

\section{Proof of Theorem 1}\label{sec:appndx_thm_proof}
Letting the number of samples tend to infinity in
\eqref{eq:cov_p_def} yields
\begin{equation}
\lim_{N\rightarrow\infty}\mathcal{C}_{\widehat{\boldsymbol{p}}}=
\lim_{N\rightarrow\infty}E\{\widehat{\boldsymbol{p}}\widehat{\boldsymbol{p}}^T\}
- \lim_{N\rightarrow\infty} E\{\widehat{\boldsymbol{p}}\}
{E\{\widehat{\boldsymbol{p}}\}}^T.\label{eq:cov_p_asym}
\end{equation}
Since the correlogram for undersampled data estimator is
asymptotically unbiased, we have
\begin{equation}
\lim_{N\rightarrow\infty} E\{\widehat{\boldsymbol{p}}\} =
\boldsymbol{p} =
\frac{\sigma^2}{W}\boldsymbol{1}_L.\label{eq:lim_p_hat}
\end{equation}
From \eqref{eq:EppT}, we obtain
\begin{eqnarray}
&&\hspace{-8mm}\lim_{N\rightarrow\infty}E\{\widehat{\boldsymbol{p}}\widehat{\boldsymbol{p}}^T\}=\nonumber\\
&&\left(\frac{L}{W}\right)^2(\breve{\boldsymbol{\Psi}}^T\breve{\boldsymbol{\Psi}})^{-1}\breve{\boldsymbol{\Psi}}^T\left(\lim_{N\rightarrow\infty}\boldsymbol{U}\right)
\breve{\boldsymbol{\Psi}}(\breve{\boldsymbol{\Psi}}^T\breve{\boldsymbol{\Psi}})^{-1}.\label{eq:limEppT}
\end{eqnarray}
Recall that all the off-diagonal elements of $\boldsymbol{U}$ are
zeros, and the first diagonal element of $\boldsymbol{U}$ is given
by \eqref{eq:Ukk3}. Letting the number of samples tend to infinity
in \eqref{eq:Ukk3}, we obtain
\begin{equation}
\lim_{N\rightarrow\infty}E\{[\boldsymbol{U}]_{1,1}\}=\sigma^4.\label{eq:limEU11}
\end{equation}
The $(Q+1)$-th element of $\boldsymbol{U}$ is zero, and if the
number of samples tend to infinity in \eqref{eq:Ukk2},
$\lim_{N\rightarrow\infty}[\boldsymbol{U}]_{k,k}=0$. Therefore, all
the elements of $\lim_{N\rightarrow\infty}\boldsymbol{U}$ are equal
to zero except for its first diagonal element which is equal to
$\sigma^4$.

In order to further simplify \eqref{eq:limEppT}, only the elements
of the first column of
$(\breve{\boldsymbol{\Psi}}^T\breve{\boldsymbol{\Psi}})^{-1}\breve{\boldsymbol{\Psi}}^T$
are required. We have shown in Appendix \ref{sec:appndx_bias} that
these elements are all equal to $1/L$. Therefore, \eqref{eq:limEppT}
can be simplified to
\begin{equation}
\lim_{N\rightarrow\infty}E\{\widehat{\boldsymbol{p}}\widehat{\boldsymbol{p}}^T\}
= \left(\frac{L}{W}\right)^2 \left(\frac{\sigma^4}{L^2}\right)
\boldsymbol{1_{LL}} = \left(\frac{\sigma^4}{W^2}\right)
\boldsymbol{1}_{LL}~\label{eq:limEppT2}
\end{equation}
where $\boldsymbol{1}_{LL}$ is an $L\times L$ matrix with all its
elements equal to $1$. It follows from \eqref{eq:cov_p_asym},
\eqref{eq:lim_p_hat}, and \eqref{eq:limEppT2} that
\begin{equation}
\lim_{N\rightarrow\infty}\mathcal{C}_{\widehat{\boldsymbol{p}}}=0.
\end{equation}
In other words, the variance of the correlogram for undersampled
data tends to zero as the number of samples goes to infinity, which
proves the consistency of the estimator. Moreover, all the elements
of $\mathcal{C}_{\widehat{\boldsymbol{p}}}$ tend to zero, which
implies that the estimations made for different spectral segments
are asymptotically uncorrelated.

\ifCLASSOPTIONcaptionsoff
  \newpage
\fi

\newpage

\end{document}